\documentclass[aps,jmp, amsmath,amssymb,reprint]{revtex4-1}
\usepackage{graphicx}% Include figure files
\usepackage{dcolumn}% Align table columns on decimal point
\usepackage{bm}% bold math
\usepackage{float}
\usepackage{epstopdf}

\newcommand{\ket}[1]{\left| #1 \right>}
\newcommand{\bra}[1]{\left< #1 \right|}

\newcommand{\Fig}[1]{Fig.\,\ref{#1}}

\begin{document}

%\preprint{AIP/123-QED}
\title{Precise simulation of single-hole spin control in quantum dots }% Force line breaks with \\
%\thanks{Footnote to title of article.}
\author{YuanDong Wang}
\affiliation{ Department of Physics, Renmin University of China, Beijing 100872, China}
\author{JingHan Ni}
\affiliation{ Department of Physics, Renmin University of China, Beijing 100872, China}
\author{JianHua Wei}
\email{wjh@ruc.edu.cn}
\affiliation{ Department of Physics, Renmin University of China, Beijing 100872, China}

\begin{abstract}
 The precise simulation of the initialization, control, and read-out of a single-hole spin qubit is investigated by accurately solving the extended Anderson impurity model in the real time domain with the hierarchical equations of motion approach. The initialization is realized by the ionization of an exciton with high fidelity. Then, a SU(2) control is achieved via the combination of Larmor procession of the hole spin in Voigt geometry magnetic field and rotation about the optical axis with a geometric phase induced by a picosecond laser pulse. Finally, the read-out of the qubit is implemented through photocurrent recording. Our theory not only reproduces the recent experimental results with one set of internal parameters, but also predicts a maximal fidelity by adjusting the dot-electrode coupling strength.
\end{abstract}
\maketitle

The trapped single spin in quantum dots (QDs) is a promising qubit which can be optically controlled  within picosecond
scale \cite{brunner2009coherent,press2008complete,de2011ultrafast}.
Therefore, it is  a good candidate for integrated circuit in quantum information processing(QIP)
with the mature processing technology of semiconductors. In literatures, many achievements have been made in
electron spin qubit in QDs \cite{press2008complete, vamivakas2010observation, delteil2014observation}.
However, the non-Markovian hyperfine interaction induces decoherence and drops the fidelity of electron spin
control \cite{vink2009locking}. The valence band holes possess {\it p}-type  wave function that leaves small residual
dipolar interaction\cite{fischer2008spin}, which highly suppresses contact hyperfine interaction and induces longer lifetimes
of hole spin than that of electron spin \cite{fallahi2010measurement,chekhovich2011direct}.
Recently, many experiments on single-hole spin qubit have been preformed, including initialization,
coherent control  and read-out \cite{ardelt2015controlled,de2011ultrafast,gerardot2008optical,godden2012coherent}.

For the realistic application of hole spin qubit, the high fidelity during the initializing process is the key requirement in any QIP protocol. Among the present methods of initialization, the ionization of an exciton has distinct advantages, which achieved  the fidelity of  $98.5\%$ \cite{ardelt2015controlled}. By reducing fine-structure splitting or applying a magnetic field  parallel to growth direction, the fidelity high to $99\%$ is
 reachable \cite{godden2010fast,brash2015high}. Moreover, it is fast (in $p$s) enough to meet the requirement that the initialization time should $10^{-4}$ order smaller than decoherence time. As a comparison, the optical pumping \cite{press2008complete}, one of other methods, can only reach the fidelity of $95\%$ in the time scale of $\mu$s due to the several loops to prepare a polarized spin state \cite{heiss2008charge}.

Whereas above experimental investigations have been actively performed, theoretical works are not sufficient so far,
and  the rate equation  is the commonly used method to simulate the hole spin manipulation \cite{breuer2002theory}.
We comment that the rate equation is not accurate enough basing on the following two facts.  Firstly, in the QDs-based hole spin
qubit, the QD directly couple to metal leads (electrodes) which inevitably impacts on the qubit.
Therefore, what we deal with is a typical quantum open system with infinity degree of freedoms of the total density matrix,
while the rate equation only concerns the diagonal terms of the reduced density matrix and treats the dot-electrode couplings
by low-order perturbation schemes. Secondly, the hole spin qubit system is a typical strongly correlated open system with degree
of freedoms of the electron-electron ($e-e$), hole-hole ($h-h$) and electron-hole ($e-h$) interactions, while the rate equation
either neglects this important interaction or treats it in the single electron level.

Obviously, for the theoretical study on  hole spin qubit,  a non-perturbative approach is highly
required to deal with the basic quantum model involving different Coulumb interactions.  The  hierarchical equations of
motion (HEOM) approach  we used in the present work can meet this requirement, which  nonperturbatively resolves the combined
effects of dot-electrode dissipation, Coulumb interactions, and non-Markovian memory \cite{jin2008exact,li2012hierarchical,ye2016heom}.
In this paper, we start from the extended Anderson impurity model to describe the hole spin qubit, fully considering the Coulumb
interactions interaction and the dot-electrode couplings. We deal with this quantum model non-perturbatively in the real time
domain to precisely simulate the single-hole spin manipulation.

\begin{figure}[]
\includegraphics[width=3.1in]{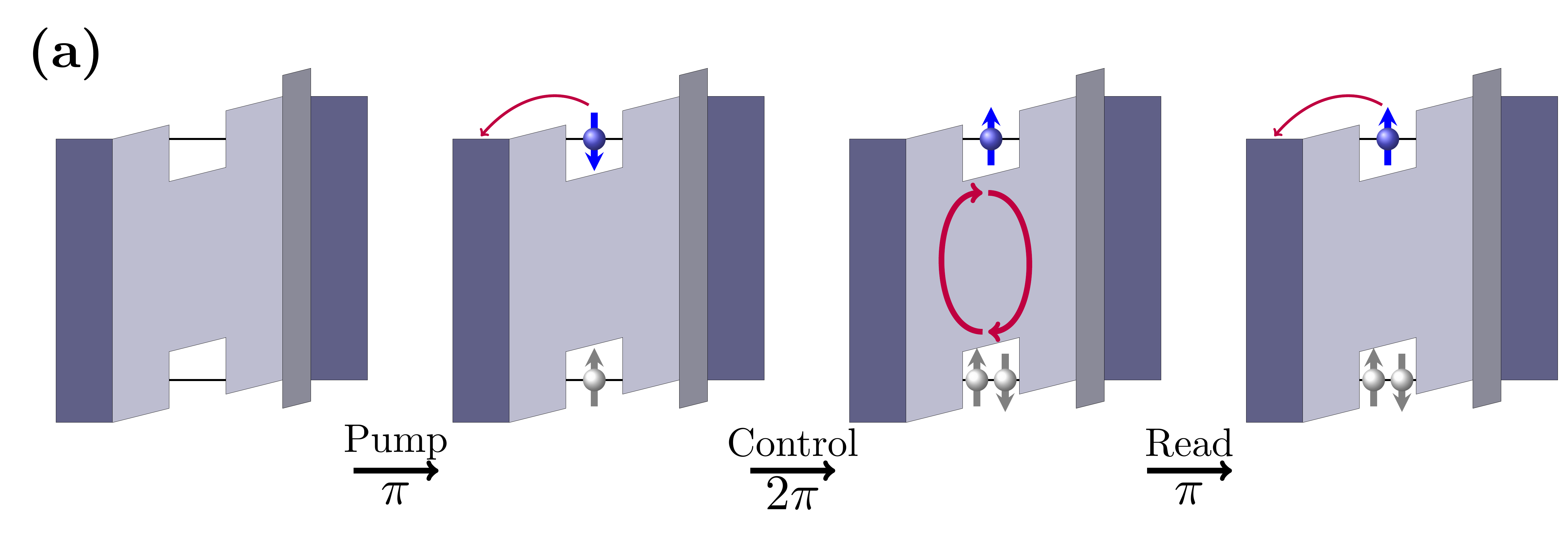}
\includegraphics[width=3.1in]{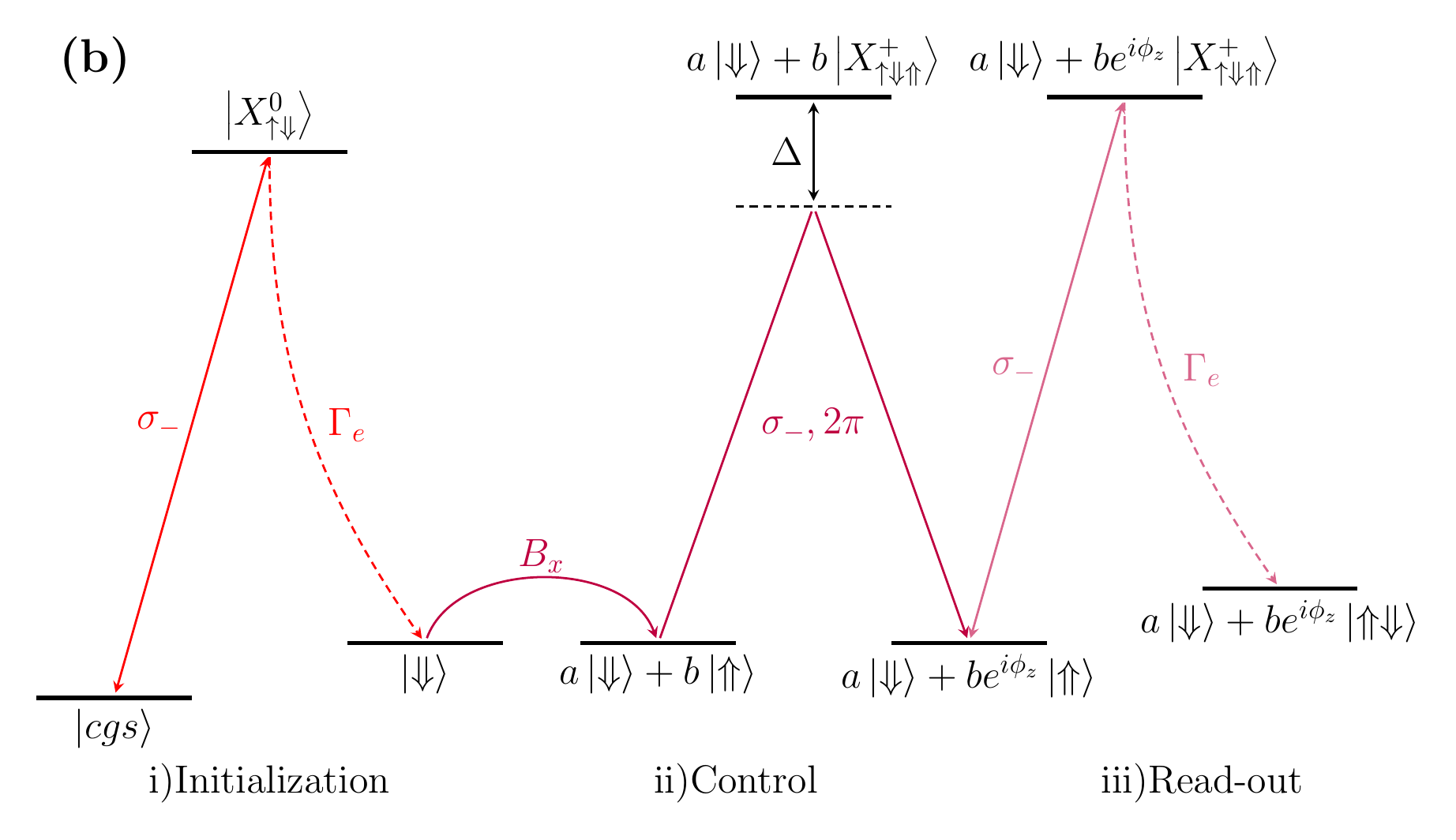}
\caption{(Color online). (a) Schematic illustration of the single-hole spin control process.
(b) Initialization, coherent control and read-out of a single-hole spin,
where the full arrows represent the optical excitation
process and the dash arrows indicate the transitions due to tunneling.}\label{fig1}
\end{figure}

In what follows, via HEOM approach, the whole process of QIP including initialization, coherent control and read-out will be
precisely simulated.  As will be demonstrated,  our theory not only reproduces the recent experimental results
in Ref.~\cite{godden2012coherent}  with one set of internal parameters, but also predicts a maximal fidelity by adjusting
the dot-electrode coupling strength.
The complete process of single hole-spin initialization, coherent control and read-out process are sketched in \Fig{fig1}(a) and (b).
In order to initialize a single-hole spin ,  a $\sigma_{+}$ resonant circularly polarized pulse with
pulse area of $\pi$  is used  which creates a $e-h$ pair that driving the ground state $\ket{cgs}$ into a neutral exciton
state $\ket{X^{0}_{\uparrow\Downarrow}}$. Due to the much larger effective mass of holes, the hybridization strength of them
is much smaller than that of electrons, i.e. $ \Delta_{H}\ll\Delta_{E}$.
As a consequence, electrons in conduction level tunnels into electrode in a $2\sim 3$ orders of time magnitude faster than holes,
which turns $\ket{X^{0}_{\uparrow\Downarrow}}$ into single-hole spin state $\ket{\Downarrow}$ quickly. Note that the operation is
confined to photocurrent region, we omit the radiative recombination of
$e-h$ pair. Then, the applied in-plane magnetic field $B_{x}$ drives the single-hole spin to precess along $x$ axis, which
preforms a U(1) operation. In order to realize a SU(2) operation in Bloch sphere, a geometric
phase approach is adopted as proposed in Ref.\cite{economou2007theory}. That with a geometric pulse of a sech envelope,
the hole spin undergoes a cycle from single hole state to positive trion state and back to single hole state acquired
an rotation angle $\phi_{z}$.
After the control sequence, a detection
circular polarized pulse $\sigma_{+}$ is applied, which partially excites the single-hole spin state to positive trion state,
accompanied with a photocurrent proportional to the spin-down component of the hole spin state, for which the qubit read-out is
achieved via photocurrent
detection technique \cite{mar2013high}.
\begin{figure*}[]
\includegraphics[width=7.0in]{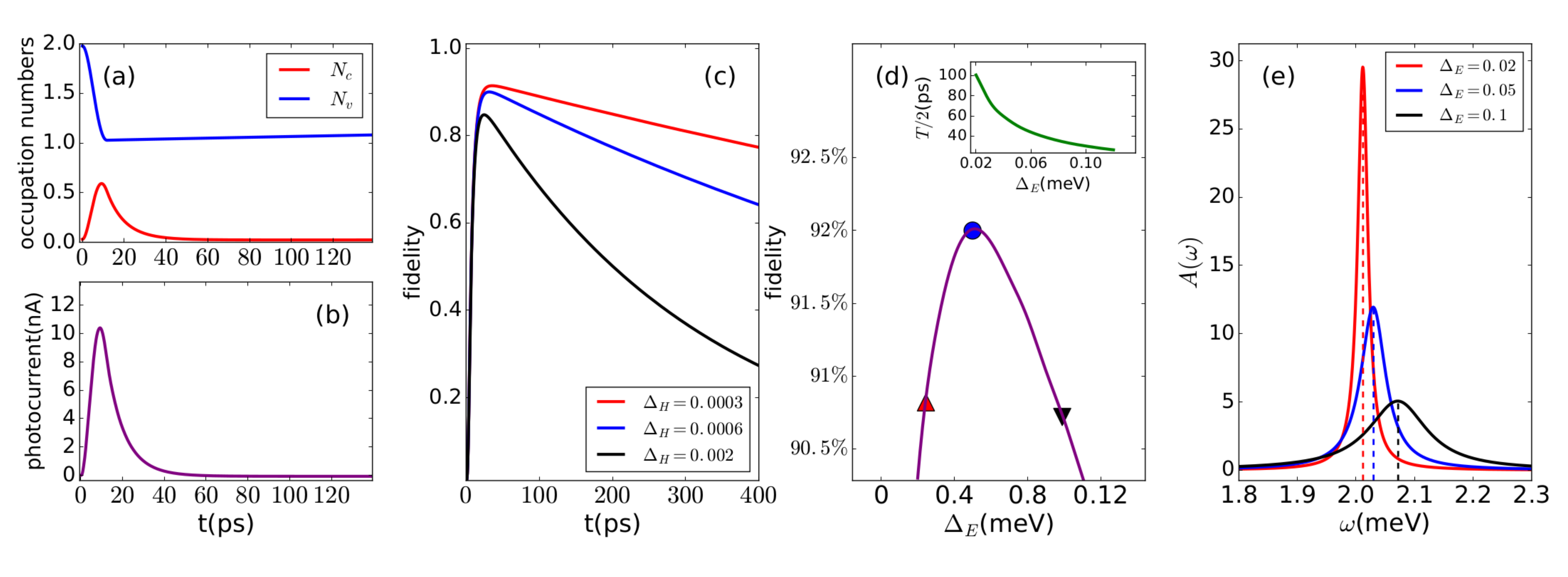}
\caption{(Color online).Initialization process, with a $\sigma_{+}$ circular polarized pulse
tuned on resonance pules applied, for which the pules area is $\pi$ and $0\le t\le 10$ ps duration. (a) The time evolution of occupation numbers of $c$-level and $v$-level.
(b) The real time photocurrent, with the maximum appeared at the end of the pulse ($t\sim 10$ ps). (c)
Fidelity as a function of time $t$ which contains the excitation-tunneling process,
with increasing the hybridization strength of hole. (d) Fidelity $F$ as a function of hybridization strength of $\Delta_{E}$, at which the maximum fidelity occurs at $\Delta_{E}\simeq 0.5$.
The inset shows the half of Rabi cycle $T/2$  as a function of $\Delta_{E}$.
(e) Density of spectral function of electrons with different electron-electrode hybridization strength. }\label{fig2}
\end{figure*}

By reference of experimental structures, our  single-hole spin qubit consists of a QD connecting to two electrodes,
which can be described by an extended Anderson impurity Hamiltonian with  $e-e$ , $h-h$ and $e-h$ interactions considered.
The total Hamiltonian is written as
\begin{equation}\label{HT}
  H=H_{c}+H_{v}+H_{c-v}+H_{opt}+H_{res}+H_{dot-res},
\end{equation}
where $H_{c}$ and $H_{v}$ describe the conduction and valence level with $e-e$  and $h-h$ interactions respectively
\begin{align}
&H_{c}=\sum_{\mu}\epsilon_{c}\hat{n}_{c\mu}+U_{c}\hat{n}_{c\uparrow}\hat{n}_{c\downarrow},\\
&H_{v}=\sum_{\mu}\epsilon_{v}\hat{n}_{v\mu}+U_{v}(1-\hat{n}_{v\uparrow})(1-\hat{n}_{v\downarrow}).
\end{align}
In above equations, $\hat{n}_{c\mu}=\hat{a}_{c\mu}^{\dagger}\hat{a}_{c\mu}$ and $\hat{n}_{c\mu}=\hat{a}_{c\mu}^{\dagger}\hat{a}_{c\mu}$,
where $\hat{a}_{c\mu}(\hat{a}_{c\mu}^{\dagger})$ annihilates(creates) a conduction level of spin $\mu$, and similar to the valence level.
 $U_{c}(U_{v})$ is the Coulomb repulsion energies if the $c$-level ($v$-level) is double (zero) occupied. The term
 \begin{equation}\label{Hcv}
H_{c-v}=-\sum_{\mu,\mu^{'}}U_{exc}\hat{n}_{c\mu}(1-\hat{n}_{v\mu^{'}})
 \end{equation}
accounts for the Coulomb attraction energies between the $e-h$ pair.
$H_{opt}$ denotes the interaction of the control field on QD,  whose explicit expression will be specified later.
The electrodes are modeled by non-interaction electrons
 \begin{equation}\label{Hres}
H_{res}=\sum_{\alpha k\mu}(\epsilon_{\alpha k}+\mu_{\alpha})\hat{d}_{\alpha k\mu}^{\dagger}\hat{d}_{\alpha k\mu},
 \end{equation}
 where $\hat{d}_{\alpha k\mu}(\hat{d}_{\alpha k\mu}^{\dagger})$ denotes the creation(annihilation) operator of electron in the specified $\alpha$-electrode spin-orbital
 state $\ket{k}$ of energy $\epsilon_{\alpha k}$. Nonequilibrium chemical potential $\mu_{\alpha}$ with $\alpha=L,R$ will arise in the presence of bias of voltage. The zero-energy point is
 set to be at the equilibrium chemical potential $\mu_{\alpha}^{eq}=0$.   The coupling between the dot and the electrode is described by
 \begin{equation}\label{Hdr}
H_{dot-res}=\sum_{\alpha k\mu}(t_{c\alpha k}
 \hat{a}_{c\mu}^{\dagger}\hat{d}_{\alpha k\mu}+t_{v\alpha k}\hat{a}_{v\mu}^{\dagger}\hat{d}_{\alpha k\mu}+H.c.),
 \end{equation}
  It should be noted that due to the large effective mass of holes, transfer matrix element for conduction level $t_{v\alpha k}$
 is assumed to be much smaller than that of valence level $t_{c\alpha k}$.
  In the HEOM theory, the influence of electron  reservoirs on the dot acts through the hybridization functions with a Lorentzian form
  $\Delta_{\mu\nu}(\omega)\equiv \sum_{\alpha}\Delta_{\alpha\mu\nu}(\omega)=\pi\sum_{\alpha k}t_{\alpha\mu k}t_{\alpha \nu k}^{*}\delta(\omega-\epsilon_{\alpha k})=
  \Delta W^{2}/[(\omega-\mu_{\alpha})^{2}+W^{2}]$, where  $W$ is the bandwidth and $\mu_{\alpha}$ is the chemical potential of lead $\alpha$.
  The details of HEOM formalism has been developed in Refs. \cite{jin2008exact, li2012hierarchical}, and the final HEOM can be cast
  into a compact form of
  \begin{align}
   \dot\rho^{(n)}_{j_1\cdots j_n} =& -\Big(i{\cal L} + \sum_{r=1}^n \gamma_{j_r}\Big)\rho^{(n)}_{j_1\cdots j_n}
     -i \sum_{j}\!
     {\cal A}_{\bar j}\, \rho^{(n+1)}_{j_1\cdots j_nj}\nonumber \\
   & -i \sum_{r=1}^{n}(-)^{n-r}\, {\cal C}_{j_r}\,
     \rho^{(n-1)}_{j_1\cdots j_{r-1}j_{r+1}\cdots j_n},
\end{align}
 where $\rho_{0}(t)=\rho(t)=tr_{res}\rho_{total}(t)$ is the reduced density matrix and ${\rho_{j_{1}...j_{n}}(t)}$ are auxiliary
 density matrices at the $n$th-tier.
  Any observable $\hat{O}$ of the
  dot system can be calculated in form of $\bar{O}=tr(\rho_{0}\hat{O})$.
  The transient current
  through the electrode $\alpha$ is determined exclusively by the first-tier auxiliary density operators as,
  \begin{equation}\label{Current}
   I_{\alpha}(t)=e\frac{i}{\hbar^{2}}\sum_{i\mu}tr_{s}\{\rho_{\alpha\mu}^{\dagger}(t)\hat{a}_{i\mu}-\hat{a}_{i\mu}^{\dagger}\rho_{\alpha\mu}^{-}(t)\},
  \end{equation}
where the index $i$ sums from $c$ to $v$ that counts the contributions both of $c$- and $v$-level. To simulate experiments,
we choose the parameters in $H_{sys}$ having the same energy level structure with experiments, as schematically shown in \Fig{fig1}(a),
with $\epsilon_{c}=2$ (the energy unit is all set to meV in rest of this paper), $\epsilon_{v}=-2$,
$U_{c}=U_{v}=2$, $U_{exc}=1$,  and a reverse bias
$V=\mu_{L}-\mu_{R}=0.2$ is applied.

For initialization, a resonant pules with $\pi$ area of 10 ps is applied, with the Hamiltonian
$H_{opt}=\Omega (e^{i\omega t}c_{c\uparrow}^{\dagger}c_{v\downarrow}+h.c.)$,
where $\omega=3$, $\Omega=0.1$, $\Delta_{E}=0.05$ and $\Delta_{H}=0.0003$.
The time evolution of occupation numbers of $c$- and $v$-level  is presented in \Fig{fig2}(a).
As shown in the figure, the maximum of the electron numbers in $c$-level is around 0.62 rather than unity,
 which is induced by the fast tunneling electron tunneling of the
neutral exciton $\ket{X_{0}}$.  It will lead to the intensity
damping of the Rabi oscillation with increasing electron-electrodes hybridization strength $\Delta_{E}$,
as a signal of tunneling-induced dephasing \cite{ardelt2015controlled}.
\begin{figure*}[]
  \centering
  \includegraphics[width=6.5in]{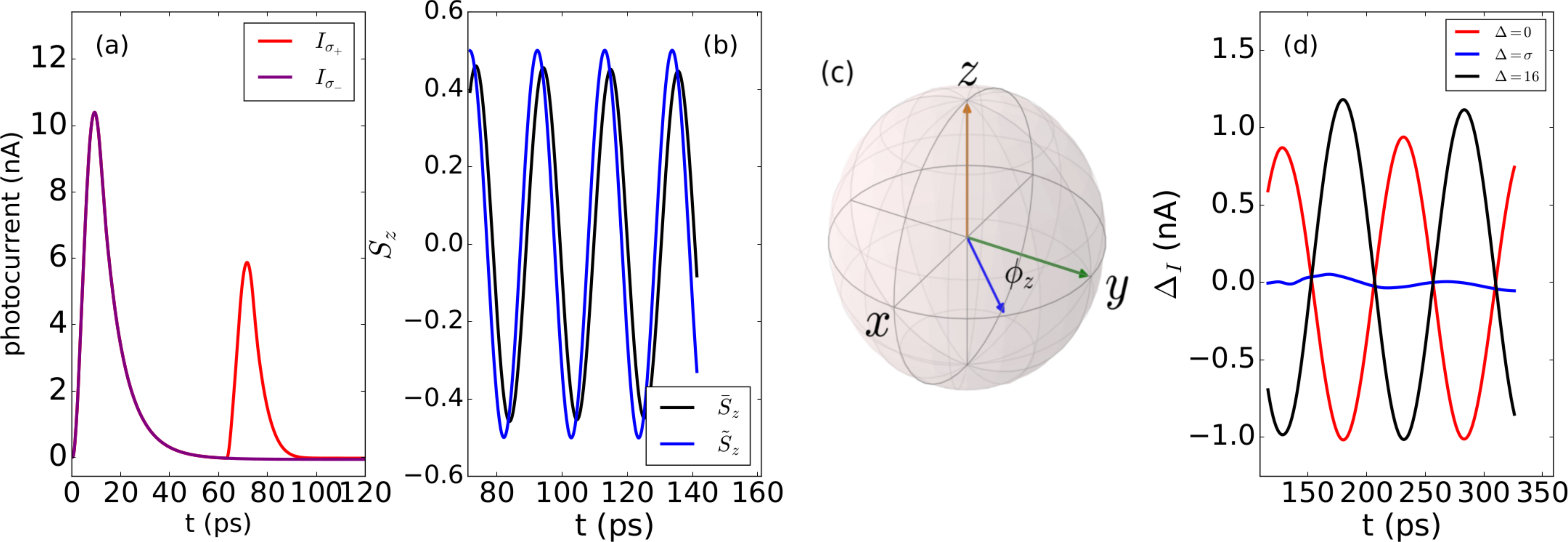}
 \caption{
 (a) The real time photocurrent $I_{\pm}$ in the presence of magnetic field $B_{x}=0.2$. Initialization pulse ends at $t=10 $ ps . With a co-circular excitation
detection pulse applied at the $\tau_{d}=\pi/f_{L}$ , no detection photocurrent produced, while with a cross-circular pulse, detection current is produced.
(b) Comparison of $z$ component of the single-hole spin calculated by $\bar{S}_{z}$ and $\tilde{S}_{z}$ .
(c) The schematic diagram of SU(2) control of a hole spin.
 the hole spin is initialization to spin up state and preforms Lamor procession about $x$ axis.
 The hole spin points along to $+y$ direction on arrival of the geometric-phase control pulse.
 (d) Photocurrent difference $\Delta_{I}=I_{+}-I_{-}$ of $\sigma_{\pm}$ detection pulse as function of precession time.
 $z$ axis rotation control is achieved via geometric-phase pulse, where the rotation angle is
 reflected via the photocurrent oscillation amplitude.
 }\label{fig3}
\end{figure*}
After the pulse applied, it can be seen that the occupation decay is exponential for electrons while approximately linear for holes at $t>10$ ps,
due to their different hybridization strength to electrodes. The initialization is achieved at $t\sim70$ ps,
where the electrons in $c$-level has totally escaped into electrodes ($N_c\sim0$) and a single hole in $v$-level has
been left ($N_v\sim1$). \Fig{fig2}(b) depicts the photocurrent along initialization process.
The peak of the photocurrent is shown at the end of the pulse ($t\sim10$ ps), where the electrons in $c$-level has maximally accumulated.
The charge-photocurrent is precisely conserved,
which is responsible for the photocurrent several orders of magnitude larger than the experimental data.

 In experiments, in order to achieve high fidelity initialization,  an AlGaAs barrier is  applied to tailor the tunneling rate of electrons and holes,
 which effectively changes the conduction and valence level-electrodes hybridization strength.
 To check this effect theoretically,  we investigate the dynamics of fidelity during the excitation-tunneling process at different $\Delta_E$ and
 $\Delta_H$. The fidelity is defined via
 $F=\overline{\bra{\psi_{in}}\hat{U}\hat{\rho}_{out}\hat{U}\ket{\psi_{in}}}$ \cite{poyatos1997complete}, which measures the distance between
 the real evolution $U$ and the target evolution $U_{t}$ to a given initial state $\ket{\psi_{in}}$. Here the initial state is set to
 $\ket{cgs}$, the target state is the single hole state
 $\ket{\Downarrow}$, which is arrived with two-step (see \Fig{fig1}(b)) process: a) optical excite  the crystal ground state $\ket{cgs}$ to
 neutral exciton $\ket{X_{0}}$,
 b) by fast electron tunneling of the neutral exciton $\ket{X_{0}}$ to single hole spin $\ket{\Downarrow}$.
 For ideal spin qubit storage with high fidelity, ultrafast $c$-level electron lifetime
 and long $v$-level hole storage against filling from electrodes are required.
 \Fig{fig2}(c)
 shows the time evolution of fidelity with different hole coupling $\Delta_H$.
 Noting that the electrons in electrodes fill the hole through the whole process, even at
 the beginning of the optical excitation. With decreasing hole coupling $\Delta_{H}$, storage time for hole increases accordingly.
 At $\Delta_{H}=0.0003$,  as much as $93\%$ of fidelity is observed.
 This fidelity dynamics can be observed via the photocurrent amplitude
 of the trion transition $X_{+}$ with a probe pulse, which reflects the population of single hole spin $\ket{\Downarrow}$.

 Ideally, one expect a high initialization fidelity with a ultrafast electron tunneling. However, when the electron hybridization strength is
 comparable to the frequency of the laser which is tuned on resonance with the neutral excitation,  it will inevitably bring damage to fidelity.
 One of these is the tunneling-induced dephasing.  Ardelt $et~al.$ extract this phenomenon
   in hole spin initialization with low  temperature  \cite{ardelt2015controlled}.  We comment that compare to rate equation method,
 the tunneling-induced dephasing is only the consequence of the increasing hybridization strength of electrons $\Delta_{E}$ in HEOM calculations, but not as a
 parameter.

The max fidelity appearing in real time scanned with $\Delta_{E}$ is shown in \Fig{fig2}(d).  With increasing $\Delta_{E}$, rather than
monotonically increase, the fidelity
experiences a maximum with $92.1\%$  at $\Delta_{E}=0.052$.
It should be noted that the couping $\Delta_{E}$ can induce a energy level shift (the details analysis will be done in \Fig{fig2}(e))
, correspondingly, in order to maximize the fidelity, the frequency $\omega$ of Rabi oscillation should be adjusted as well to
satisfy resonance condition.  The half of Rabi cycle as a function of $\Delta_{E}$ is shown in the inset of \Fig{fig2}(d).
From the beginning with $\Delta_{E}=0.02$,  as $\Delta_{E}$ increases,
 fidelity starts to grows  fast at first. This growth is dominated by the shorter electron tunneling time during which the filling of
 hole is relatively suppressed. However, continue increase $\Delta_{E}$ that exceeding $\Delta_{E}/ \Delta_{H} \simeq173$ will bring
 damage to fidelity, the fidelity drops to lower than $90\%$ with $\Delta_{E}$ exceeding $0.12$.
 In order to  explore above impact of electron-electrodes hybridization strength $\Delta_{E}$ on fidelity,
 we calculate the single particle spectral function $A(\omega)$ of $c$-level electron with $\Delta_{E}=0.02$, $0.05$ and $0.1$,  presented in \Fig{fig2}(e).
 Clearly, due to larger coupling $\Delta_{E}$, the spectral function $A(\omega)$ experiences linewidth broadening. The broadening effect
 can make the single particle level invisible to optical excitation. The ground state $\ket{cgs}$ only can be partly pumped to neutral exciton state
 $\ket{X_{0}}$  with a adjacent area around zero detuning.
 Noting that the level broadening plays a role as tunneling-induced damping, as the experiment mentioned above,
 which can be observed via the broadening of photucurrent absorption spectral \cite{ardelt2015controlled}.  The second feature of spectral function is the central position
 shift. At a small couping $\Delta_{E}=0.02$, the central position is about $\omega\simeq 2.0$, that is exactly the single particle excitation
 energy with a $c$-level electron tunneling event. When $\Delta_{E}$ increases to $0.1$, the central position moves to $2.08$, meanwhile
  the fidelity drops to $90.6\%$.

A U(1) rotation of the initialized hole spin state is achieved by applying a magnetic field $B_{x}$
perpendicular to growth direction to form the Voigt geometry.
The down-spin  state $\ket{\Downarrow}$ is a superposition of the eigenstates,  which will preform
Larmor precession about $B_{x}$ with frequency $f_{L}$ determined by the hole Zeeman splitting.
To detect the single-hole spin, a co-(cross)-circularly polarized detection pulse with $\pi$ pulse area is used to
excites the single hole to positive trion state  $\ket{X_{+}}$ after a time delay $\tau_{d}$.  %$\ket{\downarrow\Downarrow\Uparrow}(\ket{\uparrow\Downarrow\Uparrow})$,
The resonance frequency of the detection pulse is positively detuned with $U_{exc}$ compared to the initialization pulse,
due to the additional Coulomb interaction between the $e-h$ pair.
To demonstrate the photocurrent detection of a hole spin in our simulation, we depict the time evolution of photocurrent,
the detection pulse is applied at time delay $\tau_{d}=\pi/f_{L}$, for which the initialized spin down hole $\ket{\Downarrow}$
precessed to spin up state $\ket{\Uparrow}$. At this time, the excitation of co-circular detection pulse is completely suppressed
nevertheless the cross-circular pulse is optical active, which is shown in \Fig{fig3}(a).
where $\sigma_{+}$ and $\sigma_{-}$ denoting the co-circular and cross-circular excitation respectively.

In experiments, the polarization of the hole spin can be read out via photocurrent using $\tilde{S}_{z}=\frac{I_{-}-I_{+}}{I_{-}+I_{+}} $\cite{godden2012coherent},
where $I_{\pm}$ is the amplitude of photocurrent peak for the detection pulse $\sigma_{\pm}$.
Theoretically, the exact $z$ component of hole spin can be directly obtained in form of $\bar{S}_{z}$=tr$(\rho \hat{S}_{z})$, which can be used
to examine the accuracy of the photocurrent read-out process. \Fig{fig3}(b) shows the comparison between $\tilde{S}_{z}$ and
$\bar{S}_{z}$, with hole hybridization strength $\Delta_{H}=0.0003$, where the overall consistence between  $\tilde{S}_{z}$ and $\bar{S}_{z}$ can be seen.
Our results of $\tilde{S}_{z}$ is in agreement with the experiment data in Ref.\cite{godden2012coherent}, therefore, the  accuracy of single-hole spin photocurrent read-out technique
is verified by \Fig{fig3}(b). When the Rabi frequency $f_{R}$  and Zeeman splitting energy of the $B_{x}$ magnetic field is comparable, there exist a small phase difference as shown in the figure, which indicates that the time evolution of $\tilde{S}_{z}$ is slightly left behind $\bar{S}_{z}$. The phase difference results from the delay of recording photocurrent according the definition of $\tilde{S}_{z}$. As shown in \Fig{fig3}(a), the photocurrent reaches its peak value at the end of the pulse, when the hole spin has undergone a precession already.
For the parameters used here, Rabi frequency $f_{R}=0.1$ and Zeeman energy
 $B_{x}=0.2$, the phase delay is of same magnitude as half of Rabi cycle $T_{R}/2\sim 5$ps, as presented in \Fig{fig3}(b).

In order to implement a SU(2) control of the hole spin, a control pulse is needed to rotate the spin along second rotation axis.
Here we use geometric phase approach, as proposed in theories \cite{economou2006proposal, economou2007theory},
and then successfully realized in experiments \cite{godden2012coherent,kim2010fast,greilich2009ultrafast}.
The control pulse is shaped with a hyperbolic secant envelope,
$H_{opt}=\Omega {\rm sech}(\sigma t)(e^{i\omega t}c_{c\uparrow}^{\dagger}c_{v\downarrow}+h.c.)$
with fixed $\sigma=\Omega=0.2$  to guarantee no population
transferring to trion state after application of the pulse, where $\Omega$ is the Rabi frequency and $\sigma$ denotes
the bandwidth of the pulse.
The single-hole spin state acquires a phase factor
$\phi_{z}= \arctan(\frac{2\sigma \Delta}{\Delta^{2}-\sigma^{2}}) $ about $z$-axis via varying detuning
$\Delta$ from the resonance between the single hole spin state and the positive trion state \cite{economou2006proposal},
as schematically illustrated in \Fig{fig1}(b).

To be concrete, the geometric-phase control pulse is applied after a time delay $\tau=27$ ps that the hole spin pointing along +$y$ axis that the pulse
has a maximum effect on the hole spin since the rotation radius is equal to the radius of Bloch sphere, as shown in \Fig{fig3}(c).
Carries with the rotation angle $\phi_{z}$,
the hole spin continues to precess along $x$ axis under the $B_{x}$ magnetic field, for which the magnitude of geometric phase
$\phi_{z}$ determines the rotation radius along $x$ axis. Since the magnitude of the detection photocurrent is proportional to
the hole spin projection on $z$ axis, for which one can use the photocurrent difference $\Delta I=I_{-}-I_{+}$
between $\sigma_{\pm}$ detection pulse to pick up the information of the geometric rotation angle $\phi_{z}$.
The hole spin procession with different detuning $\Delta$ is shown in \Fig{fig3}(d),
that is consistent with the experiments in  Ref.\cite{godden2012coherent}, where the $\sigma_{}\pm$
detection pulses are scanned through the procession. The detunning are set as
$\Delta=0,\sigma,16$  respectively. For the detunning $\Delta=\sigma$,
the rotation angle is $\pi/2$ making the spin aligned on +$x$ axis that the precession is maximally suppressed with the
oscillation amplitude of the photocurrent nearly a constant. Nevertheless for
$\Delta=0$ and $16$ corresponding to the rotation angle $\phi=0$ and $\pi$, the precession radius reaches the maximum as well
the oscillation amplitude of photocurrent, with a difference of phase $\pi$.

In summary, by accurately solving the extended Anderson impurity model in the real time domain with the hierarchical equations of motion (HEOM) approach, we precisely simulate the whole process of single-hole spin control including initialization, SU(2) rotation, and read-out. Our theoretical results are in well agreement with the recent experimental observations, which demonstrates the feasibility and accuracy for the HEOM approach to describe the hole spin dynamics. Particularly, the influence of the hybridization strength to electrodes is fully considered, and a maximal fidelity in the initialization is predicted.

The support from the NSF of China (No.11374363) and the Research Funds of Renmin University of China
(Grant No. 11XNJ026) is gratefully appreciated.

%\subsubsection{Citations}
\bibliographystyle{aipnum4-1}
\bibliography{Refs}
\end{document}